# Parametrized Version of the Generalized Aubry-André Model

Moorad Alexanian

*Department of Physics and Physical Oceanography*
*University of North Carolina Wilmington, Wilmington, NC 28403-5606*

E-mail: alexanian@uncw.edu

**Abstract.** A recently introduced recurrence-relation ansatz applied to the Bose-Hubbard model is here used in the generalized Aubry-André model. The resulting modified Aubry-André model allows for a simple parametrization of the solutions in terms of three parameters, viz., the system energy when the quasiperiodicity amplitude $\Delta = 0$, the site $\mu$ where the particle is initially localized, and the tuning parameter $\alpha \in (-1, 1)$ that determines the regions of localized or extended states. The standard Aubry-André form corresponds to $\alpha = 0$.

## 1. Introduction

The Aubry-André (AA) model [1] is a one-dimensional crystal with incommensurate on-site energies that exhibits a localized ($\Delta/J > 2$) state to an extended ($\Delta/J < 2$) state transition, where $\Delta$ is the amplitude of the on-site energy and $J$ is the hopping energy. The AA model may be used to understand both quasicrystals and the Anderson localization metal-insulator transition in disordered systems. However, the AA model does not give rise to a mobility edge (ME), viz., an energy-dependent localization transition that demarcates localized from extended states as a function of energy. In order to generate a model that exhibits a ME, the AA model quasiperiodic site energies are generalized by introducing a tuning parameter $\alpha$, with $-1 < \alpha < 1$ that controls the shape of the potential and the distribution of site energies [2]. The generalized Aubry-André (GAA) model has been experimentally realized exhibiting an exact ME [3].

This paper is structured as follows. In Sec. 2, we present the GAA model on an infinite, one-dimensional lattice. In Sec. 3, we introduce a recurrence-relation ansatz for the term associated with the hopping term. In Sec. 4, we present our numerical calculations and compare them to experiments of nearest-neighbor, tight-binding models having a quasiperiodic site energy. Finally, Sec. 5 summarizes our results.

## 2. Generalized Aubry-André model

The GAA Hamiltonian describes a single particle moving in a disordered lattice in one dimension with hopping between nearest-neighbor sites and quasiperiodic site energies,

$$\hat{H}_{GAA} = -J \sum_n \left( \hat{c}^\dagger_{n+1} \hat{c}_n + \hat{c}^\dagger_n \hat{c}_{n+1} \right) + \sum_n \epsilon_n \hat{c}^\dagger_n \hat{c}_n, \tag{1}$$

where $\hat{c}^\dagger_n$ and $\hat{c}_n$ represent the creation and annihilation operators at site *n*, respectively, and *J* is the hopping amplitude. The quasiperiodic site energy is

$$\epsilon_n = \Delta \frac{\cos(2\pi nb + \phi)}{1 - \alpha \cos(2\pi nb + \phi)}. \quad (2)$$

The irrational number is chosen as $b = (\sqrt{5} - 1)/2$ with amplitude $\Delta$, phase $\phi$, and tuning parameter $\alpha \in (-1, 1)$. One obtains the AA model for $\alpha = 0$.

3. **Ansatz**

In order to obtain a simple parametrization of the GAA model, consider the following recurrence-relation ansatz [4] for the term associated with the hopping term of the $n$-th lattice site in (1)

$$\hat{c}_{n+1} = C(\hat{c}_n - \hat{c}_{n-1}). \quad (3)$$

and so

$$\sum_n \left(\hat{c}^\dagger_{n+1}\hat{c}_n + \hat{c}^\dagger_n \hat{c}_{n+1}\right) = B \sum_n \hat{c}^\dagger_n \hat{c}_n, \quad (4)$$

where we have considered an infinite long lattice and $B = 2C/(1 + C)$ with $B \in (-\infty, \infty)$.

The GAA model (1) is thus reduced to

$$\hat{H}_{GAA} = \sum_n (\epsilon_n - JB)\hat{c}^\dagger_n \hat{c}_n. \quad (5)$$

One can add the atomic interaction energy that corresponds to an onsite interaction, where the atoms only see each other whenever they are in the same lattice site and so

$$\hat{H}_{GAA} = \sum_n (\epsilon_n - JB)\hat{c}^\dagger_n \hat{c}_n + U \sum_n \hat{c}^\dagger_n \hat{c}_n (\hat{c}^\dagger_n \hat{c}_n - 1), \quad (6)$$

with

$$U = \frac{4\pi \hbar^2 a_s}{m} \int |\omega(x)|^4 dx, \quad (7)$$

where $a_s$ is the s-wave scattering length, $m$ is the mass of the particle and $\omega(x)$ is a Wannier function.

In order to determine the localization properties, localization is quantified by the participation ratio

$$PR = \frac{1}{\sum_n P_n^2}, \quad (8)$$

where $P_n$ is the normalized atom population at site $n$ or the inverse participation ratio

$$IPR = \sum_n P_n^2, \quad (9)$$

with $\sum_n P_n = 1$.

We consider a simple Lorentzian distribution for the normalized atom population $P_n$ at site $\mu$, viz.,

$$P_n = \frac{1}{\sum_k \frac{J/\Delta}{(k-\mu)^2 + (J/\Delta)^2}} \frac{J/\Delta}{(n-\mu)^2 + (J/\Delta)^2}, \quad (10)$$

with $0 \leq \Delta/J < \infty$, where small values of $J/\Delta$ correspond to a localized state, whereas large values of $J/\Delta$ correspond to an extended state. The atomic interaction in (6) is thus given by

$$-1 + \frac{1}{N} \leq \sum_{n=1}^{N} P_n(P_n - 1) \leq 0, \qquad (11)$$

where the upper bound corresponds to a localized state, whereas the lower bound corresponds to an extended state and $N$ is the number of sites. Accordingly, the atomic interaction gives rise to a positive (negative) contribution to the overall energy for a negative (positive) scattering length. The former corresponds to an attractive potential, whereas the latter corresponds to a repulsive potential.

The constant $B$ in (5) determines the value of $E/J$ at $\Delta/J = 0$ and is assumed to be a function of $\mu$, which is determined by the fact that the curves $E/J$ do not cross except possibly at $\Delta/J = 0$.

## 4. Numerical calculations

Consider the following expression for the energy in units of $J$,

$$E(\Delta/J, B, \alpha, \mu)/J = \sum_{k=1}^{N} \left( \frac{\Delta}{J} \frac{\cos\left(\frac{2\pi(\sqrt{5}-1)k}{2} + \phi\right)}{1 - \alpha \cos\left(\frac{2\pi(\sqrt{5}-1)k}{2} + \phi\right)} - B(\mu) \right) P_k(\mu), \qquad (12)$$

where $P_k(\mu)$ is given in (10) and we have set $U = 0$, viz., zero scattering length. Note that $E/J \to -B(\mu)$ as $\Delta/J \to 0$. The GAA model exhibits an exact ME at energy $E$ for $\alpha \neq 0$ for both $J > 0$ and $\Delta > 0$ given by [2]

$$\alpha E = 2J - \Delta. \qquad (13)$$

There ought to be no level crossing in (12) except possibly at $\Delta/J = 0$. The intersection of two energies would require

$$\sum_{k=1}^{N} \left( \frac{\Delta}{J} \frac{\cos\left(\frac{2\pi(\sqrt{5}-1)k}{2} + \phi\right)}{1 - \alpha \cos\left(\frac{2\pi(\sqrt{5}-1)k}{2} + \phi\right)} \right) (P_k(\mu) - P_k(\mu')) = B - B'. \qquad (14)$$

The right-hand-side of (14) and $P_k(\mu)$ are rational numbers, whereas the left-hand-side of (14) is an irrational number for $\Delta/J$ and $\alpha$ rational. Accordingly, our ansatz (3) preserves the non-crossing of energy eigenstates in (12) as is in (1) except possible at $\Delta/J=0$. Therefore, $\epsilon_{\mu'} > \epsilon_\mu$ implies that $B'>B$.

Consider atoms initially localized, viz., $\Delta/J \to \infty$, in the GAA model for $\alpha = -0.5$, $\phi = \pi$ and $N=201$. Fig1(a) shows the pointplot of $\epsilon_\mu/\Delta$ vs. $\mu$ from (2) of initially localized states plotted up to $\mu=201$, whereas we have only plotted up to $\mu = 15$ in Fig. 1(b). Note for $\Delta/J \gg 1$, the value of $E/J$ is actually independent of $B$. Accordingly, we consider atoms initially localized for $\Delta/J = \infty$ are slowly loaded into an eigenstate of the GAA model at a final quasiperiodicity-to-tunneling ratio $\Delta/J$. In Fig. 1(c), the plot shows $E/J$ vs. $\Delta/J$ for the range $-2 \leq B \leq 2$ for the first fifteen values of $\mu$ and choosing the value of $B$ guided by the principle of no level crossing. Clearly, we could have included all $N = 201$ initial states and thus compare our results to those of Ref. 3 more closely. However, the present results are quite consistent with those of Ref. 3. The region $E/J < -4 + 2\Delta/J$ represents extended states; whereas the region $E/J > -4 + 2\Delta/J$ represents localized states. Fig. 1(d) shows the participation ratio per unit site for $\mu = 100$ and $N = 201$, which is actually independent of the value of the tuning parameter $\alpha$. The plot shows a maximum at $\Delta/J = 0$ and is evaluated at the center of the crystal, viz. $\mu = 100$, but the result is not strongly dependent on the value of the site $\mu$. Our result shares the behavior shown in Fig. 2 (c) of Ref. 3.

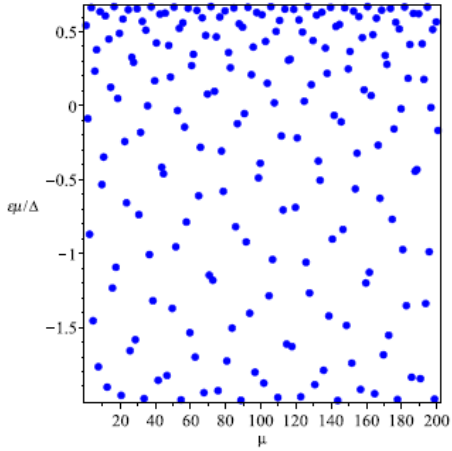

(a) Fig. 1. Pointplot of $\epsilon_\mu/\Delta$ vs. $\mu$ from (2) for $\alpha = -0.5$, $\phi = \pi$, and $N = 201$.

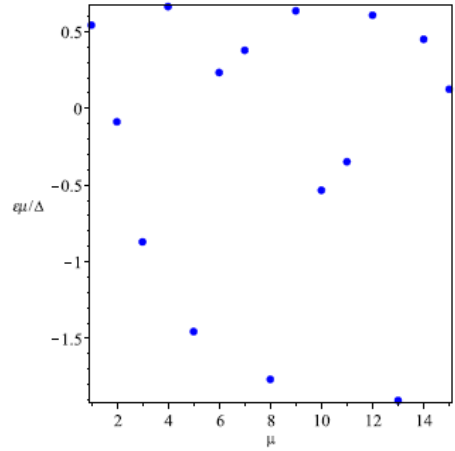

(b) Fig. 1. Same as Fig. 1(a) but including only the first fifteen values of $\mu$.

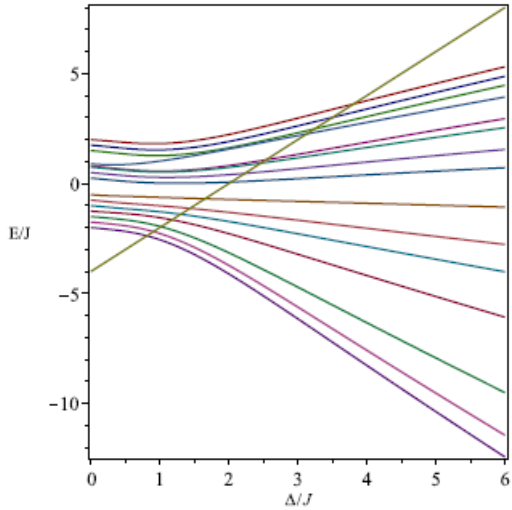

(c) Fig. 1. Eigenenergies $E/J$ vs. $\Delta/J$ of the GAA model for $\alpha = -0.5$, $\phi = \pi$ and $N = 201$ for the first fifteen sites as shown in Fig. 1(b). The ME edge is given by $E/J = -4 + 2\Delta/J$ (green).

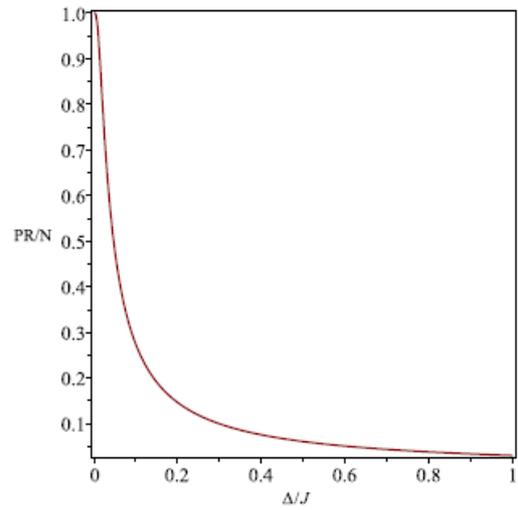

(d) Fig. 1. Participation ratio per unit site PR/N vs. $\Delta/J$ for $\mu = 100$ and $N = 201$.

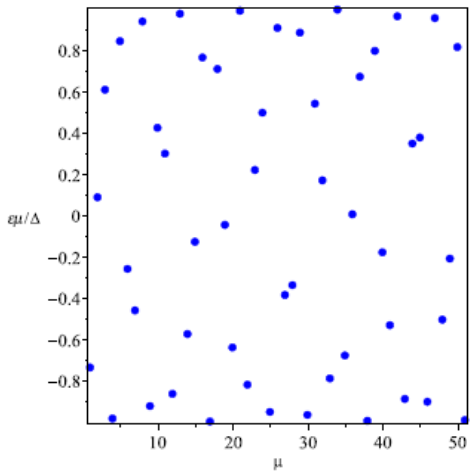

(a) Fig. 2. Pointplot of $\epsilon_\mu/\Delta$ vs. $\mu$ from (2) for $\alpha = 0$, $\phi = 0$, and $N = 51$.

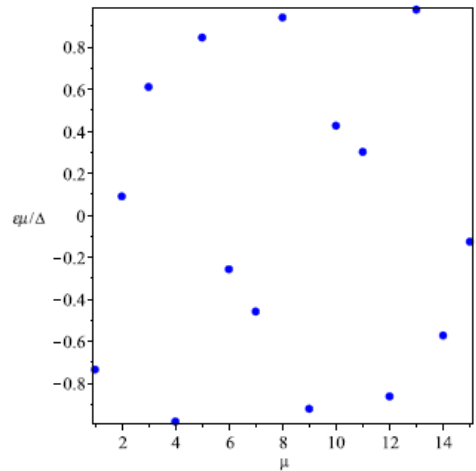

(b) Fig. 2. Same as Fig. 2(a) but including only the first fifteen values of $\mu$.

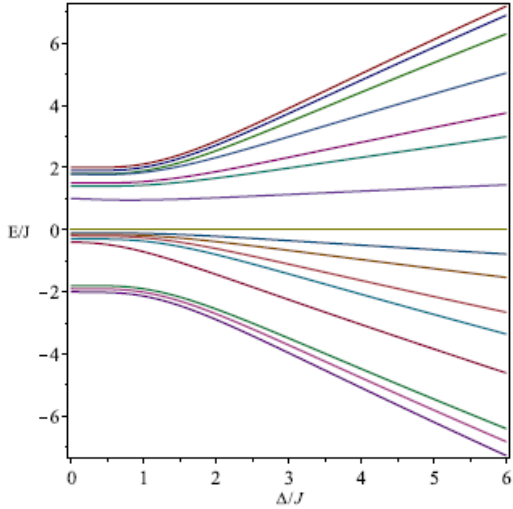

(c) Fig. 2. Eigenenergies *E/J* vs. Δ/*J* of the GAA model for *α* = 0, *ϕ* = 0 and *N* = 51 for the first fifteen sites as shown in Fig. 2(b).

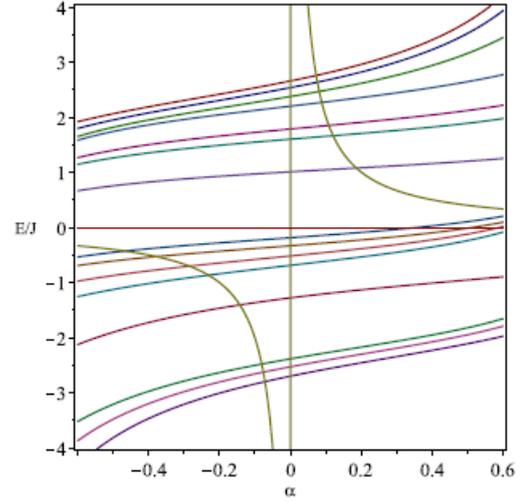

(d) Fig. 2. Eigenenergies *E/J* (12) vs. *α* of the GAA model for Δ/*J* = 1.8 and *N* = 51 for the first fifteen sites as shown in Fig. 2(b). The values of *μ* and *B* are those obtained from Fig. 2(c). The ME is given by *E/J* = .2/*α* (green).

Results in Fig. 2(a)-2(c) are obtained similarly as those shown in Figs. 1(a)-1(c), albeit, for *α*=0, *ϕ* = 0, and *N* = 51. The plot in Fig. 2(d) shows the eigenenergies just below the critical quasiperiodicity strength at Δ/*J* = 1.8. The ME, viz., *E/J* = .2/*α*, demarcates the extended from the localized states. The localized states are in the region *E(α)/J* > .2/*α* for *α* > 0 and *E(α)/J* < −.2/*α* for *α* < 0. The complementary region, −.2/*α* < *E/J* < .2/*α*, corresponds to extended states. The results qualitatively agree with those of Ref. 3.

## 5. Conclusions

We have presented a parametrized version of the generalized Aubry-André model and evaluated the eigenenergies E as a function of the quasiperiodicity amplitude Δ and separately as a function of the tuning parameter α for given value of Δ. Our results are in reasonably good agreement with the experimentally realized family of one-dimensional, nearest-neighbor, tight-binding models with quasiperiodic site energy, which model can be viewed as a generalization of the Aubry-André model.